\documentclass[doublecol]{epl2} 
\usepackage{amssymb,amstext,amsmath}
\usepackage{graphicx}
\usepackage{epsfig}
\usepackage{dcolumn}
\usepackage{bm}
\usepackage{comment}
\usepackage{subfig}

\newcommand{\noi}{\noindent}
\newcommand{\eq}[1]{\begin{align}#1\end{align}}
\newcommand{\A}{{D - z A -(1-z^2) {\bf 1}}}

\def\(({\left(}
\def\)){\right)}
\def\[[{\left[}
\def\]]{\right]}

\title{Spectral density of the non-backtracking operator}
\shorttitle{Spectral density of the non-backtracking operator} 

\author{ A. Saade\inst{1} \and F. Krzakala\inst{1,2} \and L. Zdeborov\'a\inst{3}}
\shortauthor{A. Saade \etal}

\institute{ \inst{1} Laboratoire de Physique Statistique, CNRS UMR
  8550, Universit\'e P. et M. Curie Paris 6 et \'Ecole Normale Sup\'erieure, 24,
  rue
  Lhomond, 75005 Paris, France.\\
  \inst{2} ESPCI and CNRS UMR 7083 Gulliver, 10 rue Vauquelin, Paris
  75005 France\\
  \inst{3} Institut de Physique Th\'eorique, IPhT, CEA Saclay and URA
  2306, CNRS - Orme des Merisiers 91191 Gif-sur-Yvette, France.  }


\abstract{The non-backtracking operator was recently shown to provide
  a significant improvement when used for spectral clustering of
  sparse networks. In this paper we analyze its spectral density on
  large random sparse graphs using a mapping to the correlation
  functions of a certain interacting quantum disordered system on the
  graph. On sparse, tree-like graphs, this can be solved efficiently
  by the cavity method and a belief propagation algorithm. We show
  that there exists a paramagnetic phase, leading to zero spectral
  density, that is stable outside a circle of radius $\sqrt{\rho}$,
  where $\rho$ is the leading eigenvalue of the non-backtracking
  operator. We observe a second-order phase transition at the edge of
  this circle, between a zero and a non-zero spectral density. The
  fact that this phase transition is absent in the spectral density of
  other matrices commonly used for spectral clustering provides a
  physical justification of the performances of the non-backtracking
  operator in spectral clustering.  }
\begin{document}

\maketitle

\section{Introduction}

Clustering and community detection are central tasks in the study of
social, biological, and technological networks. Sparse networks, where the average
degree of every node is a constant independent on the size of the
network, are arguably the most relevant for applications, and at the
same time the most challenging for clustering. Spectral methods are
among the most widely used for this task. They are conceptually
simply based on the computation of principal eigenvalues
and eigenvectors of an operator associated with the
network \cite{von2007tutorial}. Most commonly this operator is the
adjacency matrix, the Laplacian (symmetrized and/or normalized), the
random walk matrix, or the modularity matrix. The spectrum of these
matrices generically decomposes into a bulk of non-informative
eigenvalues, and some informative eigenvalues separated from the bulk
by a gap. The eigenvectors corresponding to the informative eigenvalues
are correlated with the cluster structure.  However, on sparse
networks, spectral clustering based on these commonly used matrices does not perform
as well as for instance methods based on Bayesian inference that can
perform well even when the tails of the bulk of the spectrum flood the
informative eigenvalue \cite{khorunzhiy2006lifshitz}.  

Recently, the author of \cite{krzakala2013spectral} proposed the
so-called non-backtracking operator for spectral clustering and
conjectured that this method is optimal: it is able to find clusters
for large random clustered networks (in the stochastic block model) as
long as it is information theoretically possible. The non-backtracking
matrix $B$, associated with an undirected graph, encodes adjacency
between directed edges. Its element $B_{i\to j,k\to l}$ is one if the edge $i\to j$ flows
into the edge $k \to l$, i.e. $j=k$ and $i\neq l$, and zero
otherwise. The authors of \cite{krzakala2013spectral} give theoretical
and numerical evidence that apart from the informative eigenvalues the
spectrum of this matrix is confined to the circle of radius the square
root of the average excess degree of the network, not presenting the
so-called Liftshitz tails \cite{khorunzhiy2006lifshitz} that spoil the
performance of spectral clustering for the other matrices mentioned
above.

In order to understand better the performance of spectral clustering
it is crucial to understand in detail the spectral properties of the
associated operators on random graphs. Analytical results for spectral
densities of sparse random graphs are largely based on the method of
replicas and cavity and were mostly developed and studied for
symmetric random matrices
\cite{edwards1976eigenvalue,rodgers1988density,semerjian2002sparse,kuhn2008spectra,rogers2008cavitySym}. The
result most relevant to the present work is that the tails of the
spectrum of the commonly studied matrices associated with random graphs
(for concreteness consider Erd\"os-R\'enyi graphs) are extended, see
e.g. \cite{khorunzhiy2006lifshitz,semerjian2002sparse,kuhn2008spectra}. On
the other hand the result of \cite{krzakala2013spectral} suggest that
the spectrum of the non-backtracking operator has no such tails.

Here, we derive the spectral density of the non-backtracking operator
for random locally tree-like graphs in the limit of large size. We use
the methods of
\cite{edwards1976eigenvalue,rodgers1988density,kuhn2008spectra,rogers2008cavitySym}
based on expressing the spectral density as the internal energy of a
disordered system with quenched disorder. In particular we use the
method applied to non-symmetric matrices as developed in
\cite{rogers2008cavity,neri2012spectra}. The corresponding disordered
system is then studied using the cavity method and the associated belief
propagation (BP) algorithm \cite{mezard2009information}.

Our main result is the discovery of a phase transition in the 
disordered system associated with the spectrum of the
non-backtracking operator which translates to the fact that the
spectral density can be different from zero only inside a circle of
radius equal to the square root of the leading eigenvalue. This is
fundamentally different from the spectral properties of the commonly
considered operators associated with a sparse random graph where the
tails of the spectrum are unbounded in the limit of large size. The
presence of this phase transition provides a physics-based explanation
of the superior performance of the non-backtracking-based spectral
clustering from \cite{krzakala2013spectral}. 

\begin{figure*}
\centering
\subfloat[Belief propagation]{%
\label{fig:BP}
\includegraphics[width=0.47\textwidth]{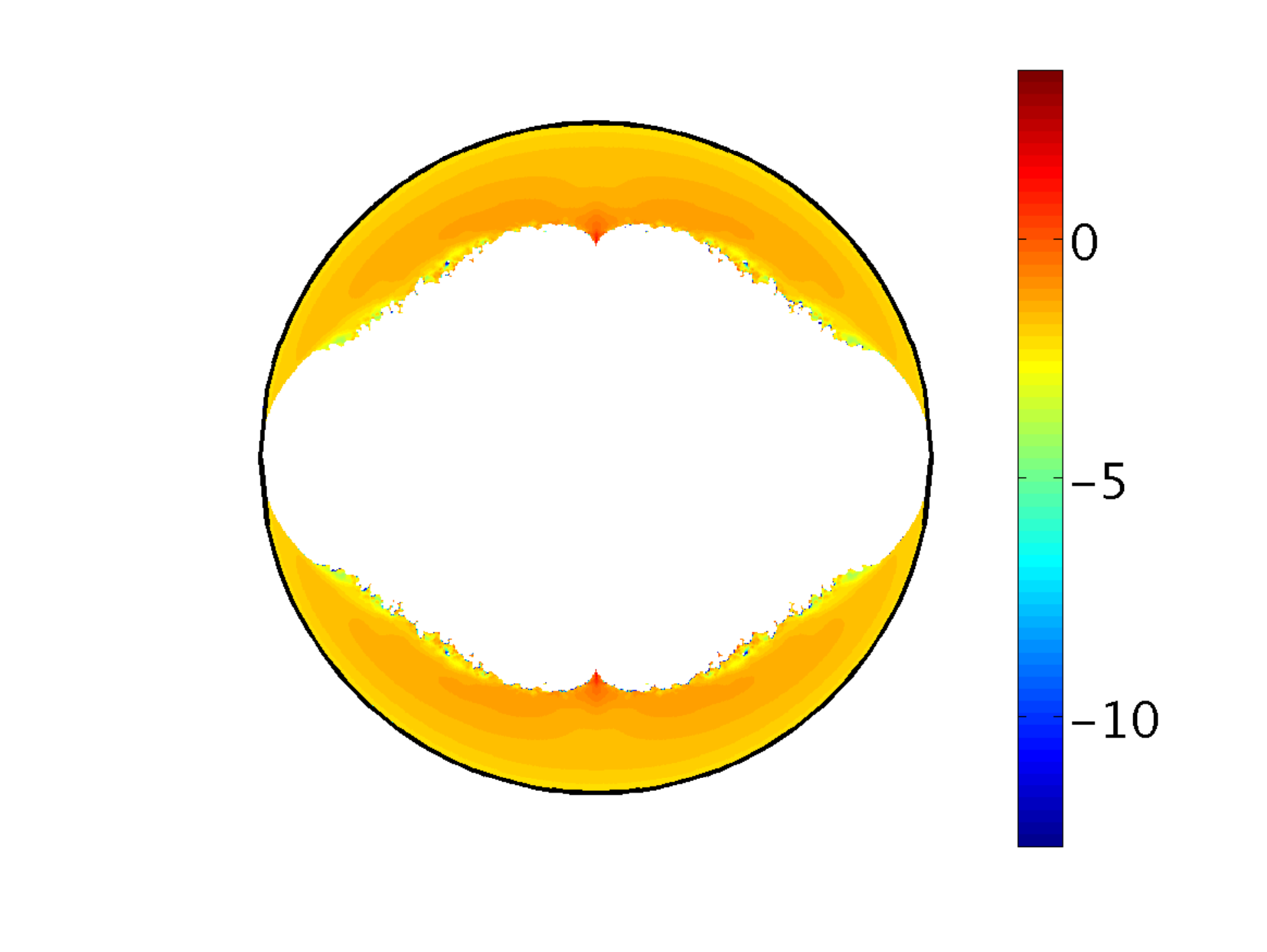}
}
\subfloat[Direct diagonalization]{%
\label{fig:DirectDiag}
  \hspace*{4pt}%
  \includegraphics[width=0.47\textwidth]{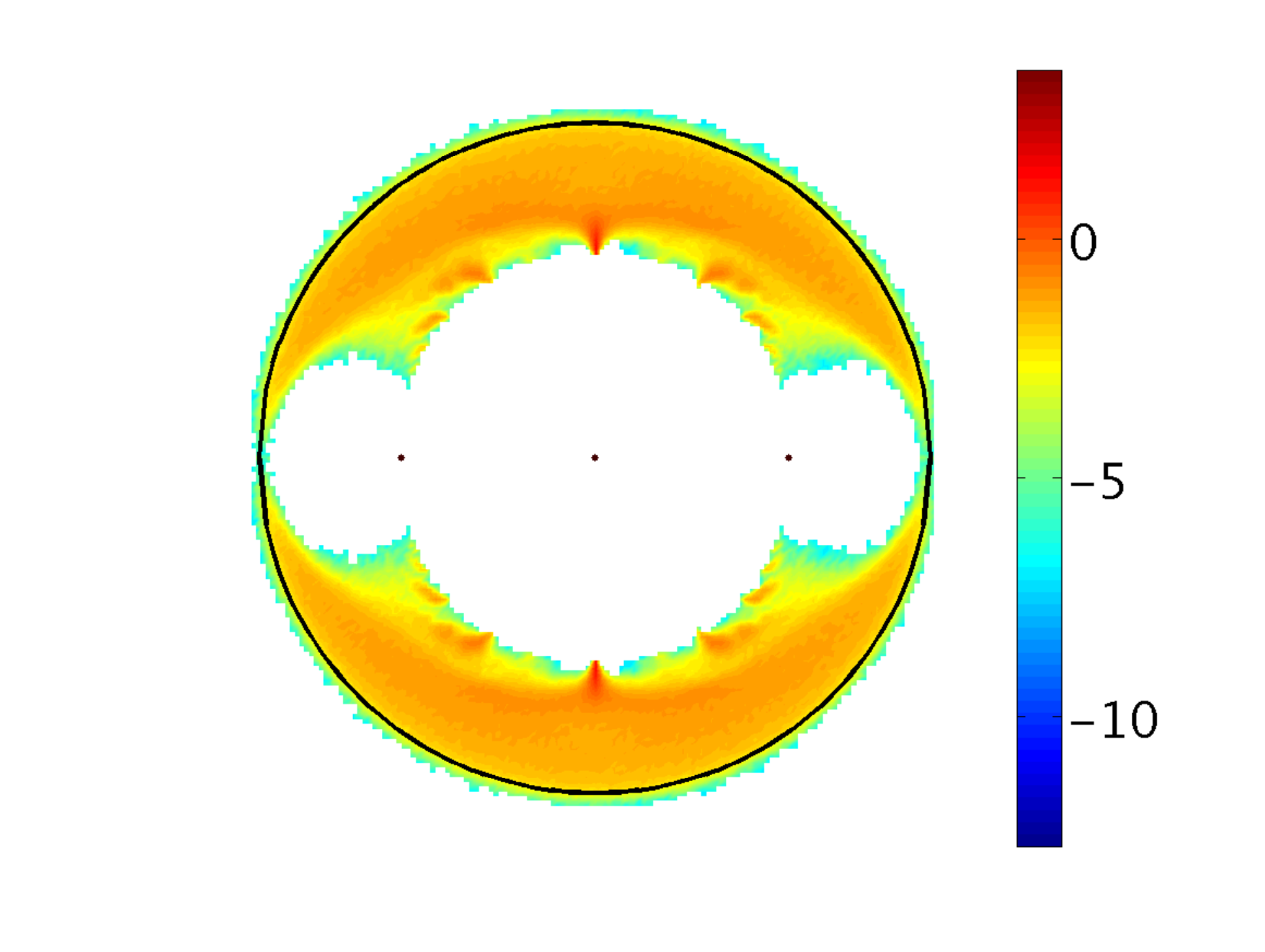}
  \hspace*{4pt}%
  }
  \caption{Spectral density of the non-backtracking matrix in
    $\ln z$-scale. Comparison between the result of belief propagation
    and direct diagonalization, on graphs of average degree
    $c=3$. Figure (b) was obtained by diagonalizing 1000 matrices of
    size $3000\times3000$. The black circle has radius
    $\sqrt{c}$. Figure (a) is the result of applying BP to a single
    graph of size 10000, at $600\times600$ different points
    $z\in\mathbb{C}$. The origins of the differences are discussed in
    the text.}
  \label{fig:spectralDensity}
  \end{figure*}

\section{Statistical physics formulation} 
To tackle the problem, following the method of
\cite{rogers2008cavity}, we map the computation of the spectral
density to a problem of statistical physics of disordered systems. 
It has been shown \cite{hashimoto1989zeta,bass1992ihara,angel2007non}
that all the eigenvalues $\lambda_i$ of $B$ that are different from $\pm 1$ are the roots of the polynomial 
\eq{
\det{\left[D - z A -(1-z^2) {\bf 1}\right]}  = \prod_i^{2N} ( z - \lambda_i
) \, .
}
This is known in graph theory as the Ihara-Bass formula.  We define
the spectral density at $z\in\mathbb{C}$ as \eq{\nu(z) = \frac{1}{2N}
  \sum_{i=1}^{2N} \delta (z -\lambda_i )\, .  } Using the complex
representation of the Dirac delta \eq{ \delta(z-\mu) = \frac
  1{\pi}\partial_{\bar{z}} (z-\mu)^{-1}\, , } where
$\partial_{\bar{z}}$ is the Wirtinger derivative, one can show that
\eq{ \nu(z)=\frac{1}{2\pi N}\partial_{\bar{z}}\partial_z
  \log\det({(\A)}^{\dag}\\\times (\A)) } whenever $z$ is not an
eigenvalue of $B$. To make this formula valid for all $z\in
\mathbb{C}$, we add an infinitesimal regularizer $\epsilon^2\bold{1}$
in the determinant, so that one can rewrite 
\eq{ \nu(z) =
  \lim_{\epsilon \to 0} \frac 1{2\pi N} \partial_{\bar{z}} {\partial}_z
  \log \det {\cal M}_{\epsilon} } 
with 
\eq{
  \nonumber &{\cal M}_{\epsilon}(z,A)=\\
  &\left( \begin{array}{cc}
      \epsilon  {\bf 1} &  i(\A) \\
      i {(\A)}^{\dag} & \epsilon {\bf 1} \end{array} \right)\, .  }
All the eigenvalues of this matrix have a positive real part
$\epsilon$, so we can use the complex Gaussian representation of the
determinant \eq{ (\det{ {\cal M}_{\epsilon}})^{-1}=\((\frac
  1{\pi}\))^{2N} \int \prod_{i}^{2N} d\psi_i d\bar{\psi}_i
  e^{-\overset{2N}{\underset{j,k}{\sum}}\bar{\psi}_j {\cal M}_{jk}
    {\psi}_k}\, .  } To take advantage of the block structure of the
kernel, we group the variables into pairs \eq{ \chi_i=
  \left( \begin{array}{c}
      \psi_i \\
      \psi_{i+N}
    \end{array} \right)
  ,\quad \forall 1\leq i\leq N\, .}
Finally, the computation of the spectral density has been mapped to a statistical physics problem 
\eq{
  \nu(z) = -\lim_{\epsilon \to 0} \frac 1{2\pi N} \partial_{\bar{z}}
  {\partial}_z \log {\cal Z}_{\epsilon}\, ,
}
where the partition function 
\eq{
  {\cal Z}_{\epsilon} = \int  d\bf{\chi} d\bf{\bar \chi} e^{-{\cal{H}}}
}
corresponds to the Hamiltonian 
\eq{\nonumber
  {\cal H}_{\epsilon}=&\sum_{i=1}^N {\chi_i}^{\dag} \left( \begin{array}{cc}
      \epsilon & i d_i - i (1-z^2) \\
      i d_i -i (1-\bar z^2)& \epsilon \end{array} \right) \chi_i \\
  +& i \sum_{i,j} {\chi_i}^{\dag} \left( \begin{array}{cc}
      0 & -zA_{ij} \\
      - \bar z {A}_{ij} & 0 \end{array} \right) \chi_j\, . \label{eq:Ham}
}
While the "Boltzmann weight" $e^{-{\cal H}_{\epsilon}}/{\cal
  Z}_{\epsilon}$ happens to be complex here, the algebraic analogy is
enough to ensure that the cavity method still works out.
Doing the derivative with respect to $z$, we can now express the
spectral density in terms of one and two-points correlation functions,
that can be computed using the cavity method 
\eq{
\label{nuVScorr}
\nu(z) = \lim_{\epsilon \to 0} \frac {i}{\pi N} \partial_{\bar{z}}
\((
z \sum_{i=1}^N 
\langle
\chi_i^{\dag} \sigma_+ \chi_i
\rangle
-
\sum_{\langle i,j\rangle}
\langle
\chi_i^{\dag} \sigma_+ \chi_j
\rangle \))\, ,
}
where the second sum is over pairs of neighbors and 
\eq{
\sigma_+=  \begin{pmatrix}
0 & 1\\
0 & 0
\end{pmatrix}\, .
}
The $\langle\rangle$ in (\ref{nuVScorr}) denotes averaging over the (complex) Boltzmann weight.

\begin{figure*}

\centering
\subfloat[Line $\text{Im}{z}=1.3$]{%
\label{fig:line13}
\includegraphics[width=0.47\textwidth]{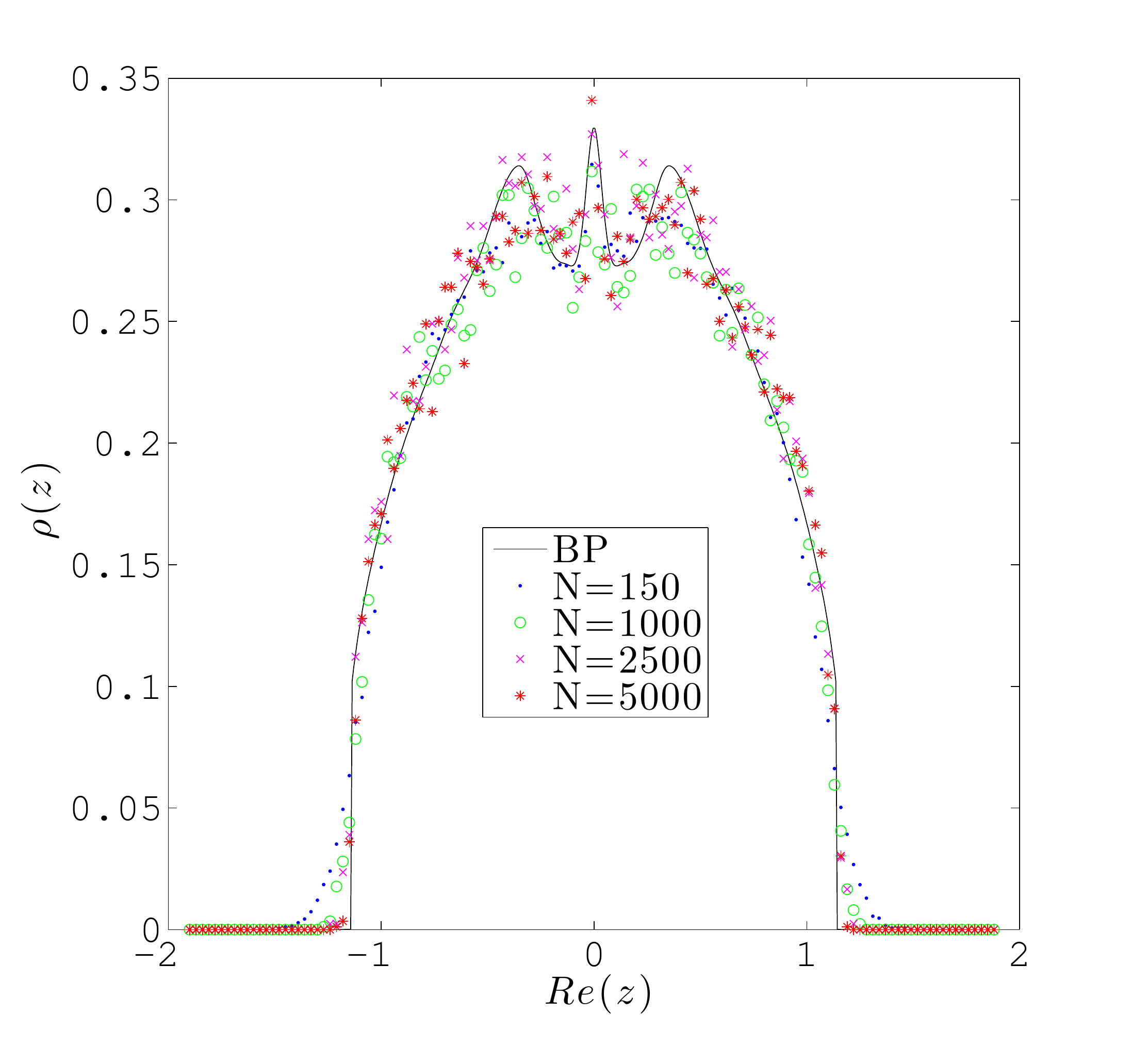}
}
\subfloat[Line $\text{Im}{z}=0.8$]{%
\label{fig:line08}
  \hspace*{4pt}%
  \includegraphics[width=0.47\textwidth]{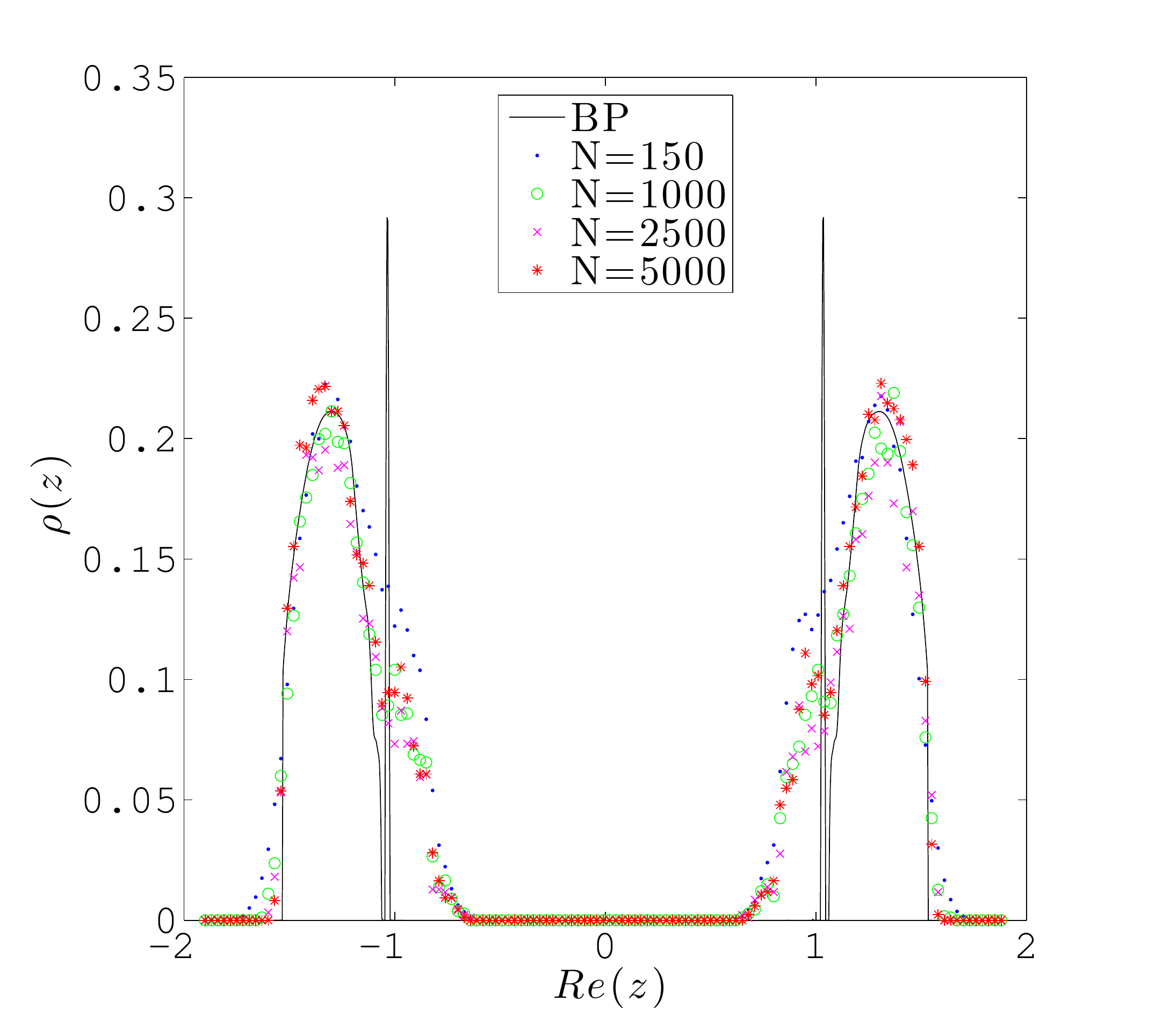}
  \hspace*{4pt}%
  }
  \caption{Slices of the spectral density along two lines: comparison
    between BP (black line), and histograms of $B$'s eigenvalues
    corresponding to graphs of size $N$, and average connectivity $c=3$. For each value of $N$
    represented, we diagonalized a number of random $B$ matrices such
    that the total number of eigenvalues obtained is equal to
    $10^6$. We then extracted those close to the line represented in
    each subfigure. The location of the peaks on the right depends on the particular instance on the graph, as shown on figure \ref{fig:sliceBP}.}
\label{fig:lines}
  \end{figure*}

\section{The cavity method} The Hamiltonian (\ref{eq:Ham}) corresponds to an
effective quantum disordered spin system that can be written as 
\eq{
{\cal H}_{\epsilon}=\sum_{i=1}^N {\cal H}_i 
+ \sum_{i<j \in \cal{G}} {\cal H}_{ij} 
}
with 
\eq{
{\cal H}_i &=\chi_i^{\dag} 
\begin{pmatrix}
\epsilon & id_i-i(1-z^2)\\
id_i-i(1-\bar z^2) & \epsilon
\end{pmatrix}
\chi_i
\\
{\cal H}_{ij} &=- i \chi_i^{\dag} 
\begin{pmatrix}
0 & z\\
\bar z & 0
\end{pmatrix}
\chi_j
-i
\chi_j^{\dag} 
\begin{pmatrix}
0 & z\\
\bar z & 0
\end{pmatrix}
\chi_i\, .
}

Denoting $P_{i\to j}(\chi_i)$ the distribution of the variable
$\chi_i$ in the absence of node $j$, and $P_i(\chi_i)$ the actual
marginal, the belief propagation recursion (exact on trees) reads \cite{mezard2009information}
\eq{
\label{cavityRecursion1}
P_{i \to j} (\chi_i) &\propto e^{-{\cal H}_i} \int 
e^{-\sum_{l    \in \partial i\backslash j} {\cal H}_{il}}\prod_{l    \in \partial
i\backslash j} P_{l \to i}  (\chi_l) d  \chi_l\, ,\\
\label{cavityRecursion2}
P_{i} (\chi_i) &\propto e^{-{\cal H}_i} \int 
e^{-\sum_{l    \in \partial i} {\cal H}_{il}}\prod_{l    \in \partial
i} P_{l \to i}  (\chi_l) d  \chi_l\, .
}
Because the Hamiltonian is quadratic, the variables $\chi_i$ are Gaussians of mean $0$, and we can parametrized 
\eq{
P_{l \to i} (\chi_l)&= \frac 1{\pi^2 \det{\Delta_{l \to i}}} e^{-
  \chi_l^{\dag} \(( \Delta_{l \to i} \))^{-1} \chi_l}\, ,\\
P_{i} (\chi_i)&= \frac 1{\pi^2 \det{\Delta_{i}}} e^{- \chi_i^{\dag}
  \(( \Delta_{i} \))^{-1} \chi_i}\, .
}
Considerations of symmetry impose the following form for the matrices $\Delta$ 
\eq{
\Delta=\begin{pmatrix}
a &   i\bar{b} \\
ib    &  a 
\end{pmatrix}\, ,
}
where $a$ is real and positive. Injecting this form in (\ref{cavityRecursion1})-(\ref{cavityRecursion2}), we find
\eq{
\label{cavity1}
&\frac{a_{i\to j}}{a_{i\to j}^2+\rvert b_{i\to
    j}\lvert^2}=\epsilon+\lvert z\rvert^2\underset{l\in\partial
  i\backslash j}{\sum} a_{l\to i}\, ,\\
\label{cavity2}
&\frac{\bar{b}_{i\to j}}{a_{i\to j}^2+\rvert b_{i\to
    j}\lvert^2}=(1-d_i-z^2)-z^2\underset{l\in\partial i\backslash
  j}{\sum}b_{l\to i}\, ,\\
\label{cavity3}
&\frac{a_{i}}{a_{i}^2+\rvert b_{i}\lvert^2}=\epsilon+\lvert
z\rvert^2\underset{l\in\partial i}{\sum} a_{l\to i}\, ,\\
\label{cavity4}
&\frac{\bar{b}_{i}}{a_{i}^2+\rvert
  b_{i}\lvert^2}=(1-d_i-z^2)-z^2\underset{l\in\partial i}{\sum}b_{l\to
  i}\, .  } \noi In the following, we take $\epsilon=0$. It only
remains to express the correlators (\ref{nuVScorr}) in terms of $a$
and $b$. Then, from (\ref{cavityRecursion2}), \eq{ \langle
  \chi_i^{\dag} \sigma_+ \chi_i \rangle = i b_i\, .  } To express the
second correlator we need the joint probability of $\chi_i$ and
$\chi_j$ where $i$ and $j$ are neighbors: \eq{
  P\((\chi_i,\chi_j\))\propto P_{j \to i} (\chi_j) P_{i \to j}
  (\chi_i)e^{{-\cal{H}}_{ij}}\, .  } Some algebra then yields \eq{
  \langle \chi_i^{\dag} \sigma_+ \chi_j \rangle =i\((-zb_{j\to
    i}b_i+\bar{z}a_{j\to i}a_i \))\, .  } Replacing in
(\ref{nuVScorr}) and using the BP recursions
(\ref{cavity1})-(\ref{cavity4}), the spectral density takes the form
\eq{
\label{densityFinal}
\nu(z)=-\frac{1}{2\pi Nz}\overset{N}{\underset{i=1}{\sum}}\((
1-d_i+z^2\))\partial_{\bar{z}}b_i \, .
}
To avoid numerical differentiation, we also compute the derivatives of these variables recursively  
\eq{
\label{cavityDer1}
\partial_{\bar{z}} a_{i\to j}=&-a_{i\to j}\((a_{i\to j}A_{i\to j}-\bar{b}_{i\to j}B_{i\to j}\))\\
&+b_{i\to j}\((a_{i\to j}C_{i\to j}+\bar{b}_{i\to j}A_{i\to
  j}\))\nonumber\, ,
\\
\label{cavityDer2}
\partial_{\bar{z}} b_{i\to j}=&-a_{i\to j}\((b_{i\to j}A_{i\to j}+a_{i\to j}B_{i\to j}\))\\
&-b_{i\to j}\((-b_{i\to j}C_{i\to j}+a_{i\to j}A_{i\to j}
\))\nonumber\, ,
\\
\label{cavityDer3}
\partial_{\bar{z}} \bar{b}_{i\to j}=&-\bar{b}_{i\to j}\((a_{i\to j}A_{i\to j}-\bar{b}_{i\to j}B_{i\to j}\))\\
&-a_{i\to j}\((a_{i\to j}C_{i\to j}+\bar{b}_{i\to j}A_{i\to j}
\))\nonumber \, ,
}

where
\eq{
\label{ABC1}
A_{i\to j}&=\underset{l\in\partial_i\backslash j}{\sum}(za_{l\to
  i}+\lvert z\rvert^2\partial_{\bar{z}}a_{l\to i})\, ,\\
\label{ABC2}
B_{i\to j}&=2\bar{z}+\underset{l\in\partial_{i}\backslash
  j}{\sum}(2\bar{z}\bar{b}_{l\to i}+\bar{z}^2\partial_{\bar{z}}
\bar{b}_{l\to i}) \, ,\\
\label{ABC3}
C_{i\to j}&=z^2\underset{l\in \partial_i\backslash
  j}{\sum}\partial_{\bar{z}}b_{l\to i}\, .
}
and similar expressions for the derivatives of the "full" variables
$a_i$ and $b_i$ in which $a_{i\to j}$ gets replaced by $a_i$ in
(\ref{cavityDer1})-(\ref{cavityDer3}), and the sums in
(\ref{ABC1})-(\ref{ABC3}) become over all neighbors of $i$.  Equations
(\ref{cavity1})-(\ref{cavity2}) and (\ref{cavityDer1})-(\ref{ABC3})
are self-consistent BP equations which, when iterated,
converge to a set of solutions $a_{i\to j}$, $b_{i\to j}$,
$\partial_{\bar{z}} a_{i\to j}$, $\partial_{\bar{z}} b_{i\to j}$,
$\partial_{\bar{z}} \bar{b}_{i\to j}$. We then compute the full
variables $a_{i}, b_{i}$, $\partial_{\bar{z}} a_{i}$,
$\partial_{\bar{z}} b_{i}$, $\partial_{\bar{z}} \bar{b}_{i}$ using
equations (\ref{cavity3})-(\ref{cavity4}) and the counterpart of
equations (\ref{cavityDer1})-(\ref{ABC3}) for the full variables. This
finally allows us to compute the spectral density using expression
(\ref{densityFinal}).

\section{The paramagnetic phase}
It is easy to see that the following assignment of the
variables is a fixed point of the belief propagation equations 
\eq{
\label{paraSolCav}
&a_{i\to j}=0      &\forall \langle i,j\rangle\, , \\
&b_{i\to j}=-\frac{1}{z^2} &\forall \langle i,j\rangle} \, 
We call this the {\it factorized} fixed point. 
The corresponding assignment of the full variables is 
\eq{
a_{i}=0   ~~~ \forall \langle i,j\rangle 
~~~~~
b_{i}=\frac{1}{1-z^2} ~~~\forall \langle i,j\rangle}
With this solution, we have $\partial_{\bar{z}}b_i=0$ for all $i$ so
that, from (\ref{densityFinal}), the spectral density is $\nu(z)=0$.
Wherever the above solution is stable, the cavity method yields a
spectral density equal to $0$. We will therefore refer to this region
as the paramagnetic phase. To study the stability of this solution, we
linearize the belief propagation equations
(\ref{cavity1})-(\ref{cavity2}) around it. Writing $a_{i\to
  j}=0+\alpha_{i\to j}^0,\ b_{i\to j} =-1/z^2+\beta_{i\to j}^0$ where
$\alpha_{i\to j}^0\in \mathbb{R},\beta_{i\to j}^0\in \mathbb{C}$ are
the initial infinitesimal perturbations, the evolution of these
perturbations when iterating the BP equations is given by the system
\eq{
\label{CavityLin}
&\alpha_{i\to j}^{t+1}=\frac{1}{\lvert
  z\rvert^2}\underset{l\in\partial i \backslash j}{\sum}\alpha_{l\to
  i}^{t}\, ,\\
&\beta_{i\to j}^{t+1}=\frac{1}{z^2}\underset{l\in\partial i \backslash
  j}{\sum}\beta_{l\to i}^{t}\, .
}

  \begin{figure}

\centering

\includegraphics[width=0.49\textwidth]{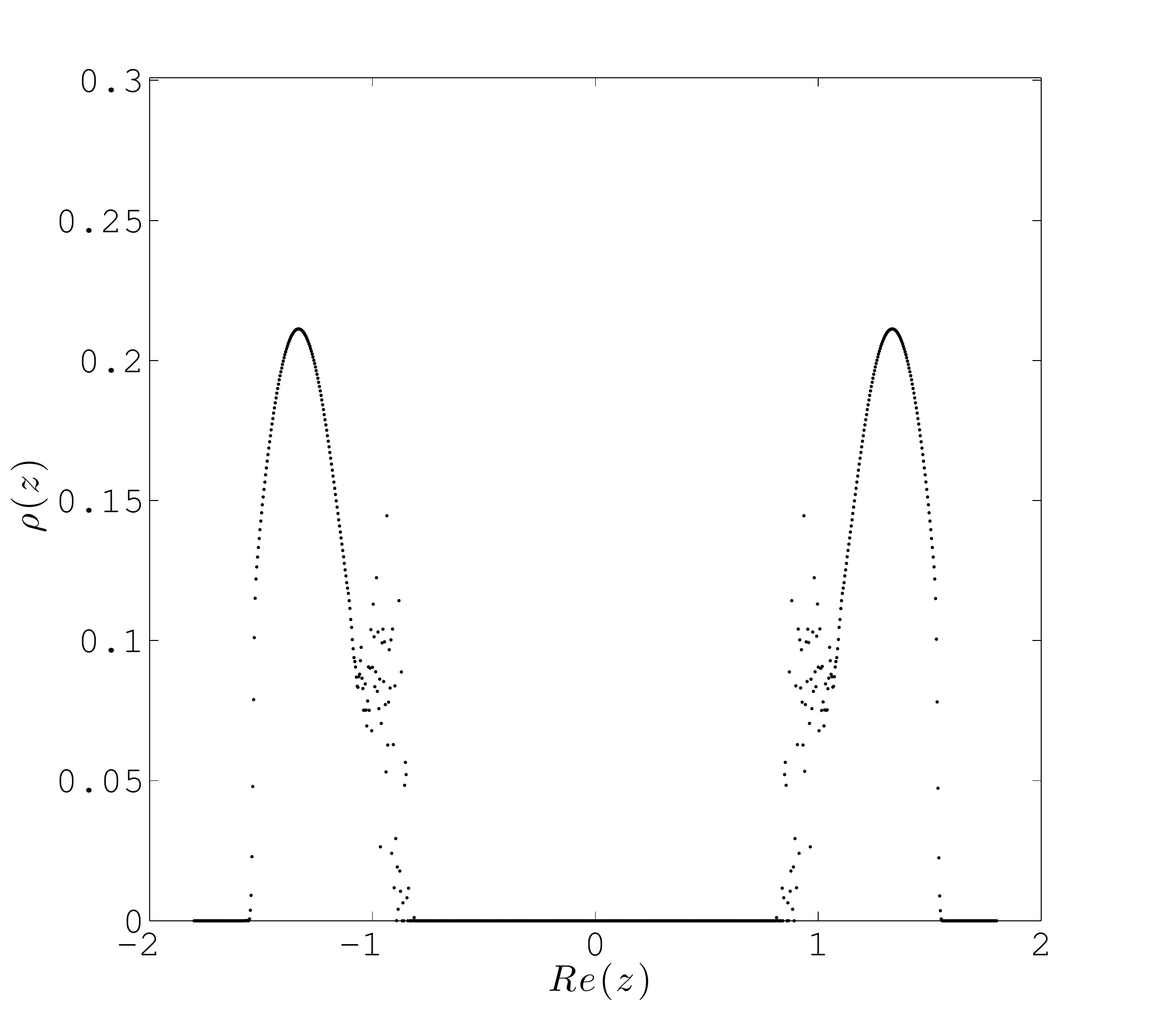}
\caption{Slice of the spectral density along the line
  $\text{Im}(z)=0.8$: average over $500$ different random graphs with
  $10 000$ nodes, and were generated with an average degree of
  $c=3$. Comparison with figure \ref{fig:line08} shows that the
  location of the peaks inside the circle depends on the instance.}
\label{fig:sliceBP}
  \end{figure}

One can rewrite this system in a matrix form using the
non-backtracking operator $B$. It was already remarked in
\cite{krzakala2013spectral}, although in a completely different
setting, that $B$ arises from the linearization of belief propagation
around a factorized fixed point. The linear relations then reads
\eq{
\label{linB}
\underline{\alpha}^{t+1}&=\frac{1}{\lvert z\rvert^2}B^{T}\underline{\alpha}^t\\
\underline{\beta}^{t+1}&=\frac{1}{z^2}B^{T}\underline{\beta}^t
}
where $\underline{\alpha},\underline{\beta}$ are two vectors of size $2M$, $M$ being the number of edges of the graph. From these equations we see that the paramagnetic solution is stable if and only if the largest eigenvalue of $B$ (in absolute value) is smaller than $\lvert z\rvert^2$. Therefore the bulk is constrained to the disk
\eq{
\label{bulk}
\lvert z\rvert\leq \sqrt{\rho({B})}
}
where we have introduced the spectral radius $\rho({B})$ of the
non-backtracking operator. Instability of the factorized belief
propagation fixed point signals a phase transition in the associated
particle system. 

The above result is valid for any graph where the cavity method
applies. We expect this to encompass at least all locally tree-like
ensembles. Eq. (\ref{bulk}) supports the heuristic used
\cite{krzakala2013spectral} on real networks to consider as
informative only the eigenvalues that lie outside of the circle of
radius $\sqrt{\rho({B})}$. For an Erd\"{o}s-R\'enyi graph $\rho(B)=c$.

\begin{figure*}

\centering

\includegraphics[width=0.93\textwidth]{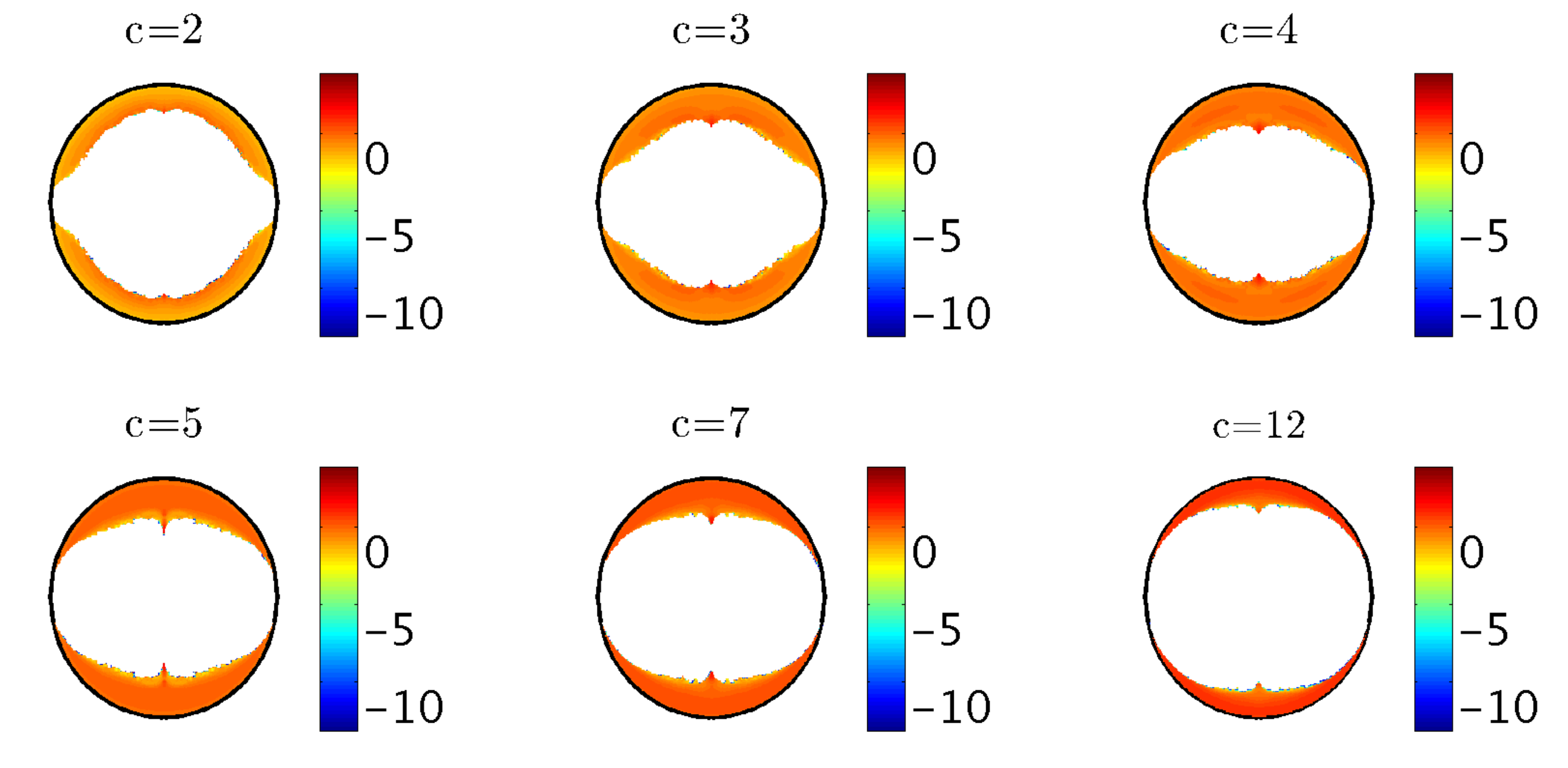}

\caption{Spectral density for different average degrees in $\ln$ z-scale. The support becomes closer and closer to the circle with bigger $c$, as expected because of the exact computation in the regular case.}
\label{fig:6conn}
\end{figure*}

The existence of a factorized fixed point, and hence of a paramagnetic
phase in which the spectral density is exactly $0$, seems to be a
special feature of the non-backtracking matrix. For instance, one can
compute the spectral density for the (symmetric) adjacency matrix $A$,
see e.g. \cite{rogers2008cavitySym}.  The Hamiltonian is then again
quadratic, and couples $N$ Gaussian variables $x_i$. The marginals of
the $x_i$ are again completely determined by their (complex) variance
$\Delta_i$, as is the spectral density which is
proportional to the average of
$\text{Im}{\Delta_i}$ over the graph. Using the same notations as
before, the BP equations read \cite{rogers2008cavitySym} \eq{
\label{cavityA}
\Delta_{i\to j}(z)=\((z-\underset{l\in\partial i\backslash j}{\sum}\Delta_{l\to i}(z) \))^{-1}
}
for which no factorized (site-independent) solution exists. 
The spectral density of the adjacency matrix instead exhibits Lifshitz
tails \cite{khorunzhiy2006lifshitz} that spoil the gap between the bulk and the eigenvalues
reminiscent of the presence of clusters. Similar results hold for the
other matrices commonly used for spectral clustering.

\section{Numerical results} We solve the belief propagation equations on a single
graph. For a given point $z\in \mathbb{C}$ in the complex plane, we
iterate (\ref{cavity1})-(\ref{cavity2}) and
(\ref{cavityDer1})-(\ref{cavityDer3}) until convergence, and output
the spectral density as given by (\ref{densityFinal})\footnote{Another
  approach, aiming at computing the spectral density in the
  thermodynamic limit is the population dynamics algorithm which we
  also implemented. It, however, suffered from lack of convergence,
  symptomized by a non-vanishing imaginary part of the spectral
  density.}. BP is found to always converge to a real-valued spectral
density. 

Figure \ref{fig:BP} shows the results of BP for a typical random graph
of size 10000, with average degree $c=3$. We expect this figure to be
a fairly close approximation of the thermodynamic limit. For
comparison, figure \ref{fig:DirectDiag} shows the spectral density as
computed by histogramming the eigenvalues of many matrices. The
discrepancies between the two figures are of two types. The first one
consists of the tails that extend beyond the black circle in the
direct diagonalization case. These represent sub-extensive
contributions to the spectral density as can be seen from figure
\ref{fig:lines}, that disappear in the thermodynamic limit, in
agreement with the prediction from BP. The second type of discrepancy
consist of the tails inside the circle in figure \ref{fig:DirectDiag}
that are absent from figure \ref{fig:BP}. As can be seen from
\ref{fig:line08}, these tails do not seem to vanish in the large $N$
limit. As supported by figure \ref{fig:sliceBP}, they seem to
originate from very localized peaks in the spectral density, that BP
fails to see because of the finite resolution of the grid used.
Finally, we show in figure \ref{fig:6conn} the spectral density
computed by BP for 6 different values of the average degree. As shown
in \cite{krzakala2013spectral}, in the regular case, the spectral
density is non-zero only on the circle of radius $\sqrt{d-1}$. For an
Erd\"{o}s-R\'enyi graph, in the large degree limit, we expect the
spectral density to be non-zero only in a region close to the circle
of radius $\sqrt{c}$, because for a Poissonian graph, the excess
degree is also Poissonian of mean $c$.

\section{Conclusion} 
The study of the spectral density of the non-backtracking operator in
the thermodynamic limit by means of the cavity method allows to
understand better its remarkable efficiency to perform spectral
clustering. A phase transition-like behavior at the boundary of the
circle provides a physical insight on why its spectral density
vanishes sharply instead of exhibiting Lifshitz tails, like other
popular choices of spectral methods.  Additionally, the
non-backtracking operator seems to have puzzling properties still
unexplained, like the ability to predict the number of clusters in a
graph from counting real eigenvalues outside of the circle of radius
$\sqrt{\rho(B)}$. This fact will be investigated in future work.

\acknowledgments
This work has been supported by the ERC under the European Union'€™s 7th
Framework Programme Grant Agreement 307087-SPARCS.

\bibliographystyle{eplbib}
\bibliography{mybib}

\end{document}